\documentclass{elsart}
\usepackage[dvips]{graphics}
\usepackage{amssymb}
\usepackage[dvips]{graphics}

\begin{document}

\title{Logarithmic Universality in Random Matrix Theory} \author K. Splittorff \ email: split@alf.nbi.dk \\ The Niels Bohr Institute \\ Blegdamsvej 17 \\ DK-2100 Copenhagen \O \\telephone: 4535325401 \\ fax: 4535325400 \\ Denmark

------------------------------------------------------------------------------------

\noindent
{\bf Abstract}
\vspace{2mm}
\small

Universality in unitary invariant random matrix ensembles with complex matrix elements is considered. We treat two general ensembles which have a determinant factor in the weight. These ensembles are relevant, e.g., for spectra of the Dirac operator in QCD. In addition to the well established universality with respect to the choice of potential, we prove that microscopic spectral correlators are unaffected when the matrix in the determinant is replaced by an expansion in powers of the matrix. We refer to this invariance as logarithmic universality. The result is used in proving that a simple random matrix model with Ginsparg-Wilson symmetry has the same microscopic spectral correlators as chiral random matrix theory.
\vspace{2mm}

\vspace{2mm}
\normalsize

------------------------------------------------------------------------------------------------------

\section{Introduction}

Random matrix theory (RMT) deals with universal spectral properties of ensemble averages of large random matrices \cite{Metha}. This universality is revealed in the microscopic limit. In the microscopic limit we consider the spectral properties of eigenvalues $\lambda$ of matrices of dimension $N$ in the limit as $N\to\infty$ with $x\equiv N\lambda$ fixed \cite{ShuVer}. 
Studies of universality in RMT can be divided into two categories. In one category the models studied have unitary invariance (see, e.g., \cite{ADMN}) while in the other this invariance is violated (see, e.g., \cite{TandT}). In this paper we consider the former category. The partition function defining the ensemble, hence, can be expresses in terms of the eigenvalues of the random matrices rather than in terms of the matrices. We consider random matrix models where the partition functions involves a general potential $V^{(\alpha)}(\lambda)$.
This general potential is a sum of a regular and a logarithmically singular part. This paper introduces and proves logarithmic universality. Logarithmic universality expresses the invariance of microscopic spectral correlators with respect to deformations of the logarithmically singular part of $V^{(\alpha)}(\lambda)$. The logaritmically singular part in the eigenvalue representation originates in the determinant of a matrix $D$.   

Two random matrix ensembles will be studied: The unitary ensemble (UE) relevant, e.g., for QCD in 3 dimensions \cite{QCD3,DNQCD3,jesperI} and the chiral unitary ensemble ($\chi$UE) relevant, e.g., for QCD in 4 dimensions \cite{ShuVer,VerZah,Verorig}. In the applications of RMT to QCD the matrix, $D$, in the determinant is analogous to the Dirac operator.   
For the $\chi$UE the logaritmic universality includes deformations which violate both the chiral symmetry condition $\{D,\gamma_5\}=0$ and the hermiticity. As an example we will show that a simple random matrix model which satisfies the Ginsparg-Wilson relation $\{D,\gamma_5\}=D\gamma_5D$ \cite{GW} shares the microscopic correlators of the $\chi$GUE.

\noindent
In the first part of this paper, we introduce and prove logarithmic universality in the unitary ensemble. In order to do this we make use of a method established by Kanzieper and Freilikher \cite{KandFWork}. In the second part, we introduce logarithmic universality in the chiral unitary ensemble and use the universality of the UE to prove the logarithmic universality of the microscopic spectral correlators of the $\chi$UE. Finally, we consider the example with the Ginsparg-Wilson symmetric model.

\section{Logarithmic universality in the unitary ensemble}

The unitary ensemble is defined by the partition function (see e.g. \cite{multicritical})
\begin{eqnarray}
\mathcal{Z}_{\rm UE} & \equiv & \int {\rm d}M {\rm det}^{2\alpha} M e^{-N{\rm Tr}V(M)} \\
         & \sim   & \int_{-\infty}^\infty \prod_{i=1}^N\left({\rm d}\lambda_i \lambda_i^{2\alpha} \exp\{-NV(\lambda_i)\}\right) |\Delta(\lambda)|^2 \ ,
\end{eqnarray}
where the $\lambda_i$ are the eigenvalues of the hermitian $N\times N$ matrices $M$ and  $V(M)$ is an expantion in even powers of $M$
\begin{equation}
V(M)=\sum_{k=1}^p\frac{g_{2k}}{2k}M^{2k}\ , \ \ g_{2k}\in \mathbb{R} \ {\rm with}\ g_{2p}>0 \ . 
\end{equation}
The Vandermonde determinant,
\begin{equation}
\Delta(\lambda)\equiv\prod_{j>i=1}^{N}(\lambda_i-\lambda_j) \ ,
\end{equation}
is the Jacobian describing the change to angular coordinates \cite{morris}. Finally, the measure of the matrix integration is the Haar measure.

The hard edge, bulk, and soft edge correlators of this ensemble have been studied in detail and have been shown \cite{ADMN} to be independent of the coefficients $g_{2k}$ in the potential $V$ in the microscopic limit.   

We now introduce a new set of deformations and show that the microscopic spectral correlators are invariant with respect to these deformations.
\vspace{2mm}

\noindent
We shall prove that {\sl the microscopic spectral correlators of the partition function
\begin{eqnarray}
\mathcal{Z}_{\rm UE} & = & \int {\rm d}M {\rm det}^{2\alpha}\left( f(M) \right)\exp\{-N{\rm Tr}V(M)\}
\label{invUZ(M)} \\
         & \sim &  \int_{-\infty}^\infty \prod_{i=1}^N\left({\rm d}\lambda_i [f(\lambda_i)]^{2\alpha} \exp\{-NV(\lambda_i)\}\right) |\Delta(\lambda)|^2 
\label{invUZ(l)}
\end{eqnarray}
are independent of the coefficients $f_k\in\mathbb{R}$ in the polynomial 
\begin{equation}
f(M)=\sum^\infty_{k=0}f_{2k+1}M^{2k+1}
\label{fexp}
\end{equation}
with $f_1\not=0$.}
\vspace{5mm}
The expansion defining $f$ is assumed to have suitable convergence properties. We denote this independence as logarithmic universality. 

In the proof of this assertion, we shall follow, and extend where necessary, the argument of Kanzieper and Freilikher \cite{KandFWork}. At present \cite{KandFWork} provides the most far-reaching proof of the range of the universality class. At the same time, this method enables us to explore the spectral properties of the matrix model defined in eqs.\,(\ref{invUZ(M)}) and (\ref{invUZ(l)}) for an arbitrary spectral range. Without loss of generality, we can choose $f_1=1$.

\subsection{Proof of logarithmic universality in the UE}

For notational simplicity, we will absorb the $f$-dependent part of the integrand in the partition function of eqn.\,(\ref{invUZ(l)}) into a general potential 
\begin{equation}
V^{(\alpha)}(\lambda)\equiv V(\lambda)+V^{(\alpha)}_{\rm sing}(\lambda)=\sum_{k=1}^p\frac{{\rm g}_{2k}}{2k}\lambda^{2k}-\frac{\alpha}{N}\log[f(\lambda)]^2 \ .
\label{genpot}
\end{equation}
\vspace{2mm}

%Since the partition function of eqn.\,(\ref{invUZ(M)}) is unitarily invariant the method of orthogonal polynomials will be applied. 
In order to determine the kernel, $K_N(\lambda,\lambda')$, from which all spectral correlators follow,
\begin{equation}
\rho(\lambda_1,\lambda_2,\ldots,\lambda_s)=\det_{1\leq a,b\leq s}K_N(\lambda_a,\lambda_b) \ ,
\label{rhofraK}
\end{equation}
it is sufficient to find the set of polynomials $P_n(\lambda)$ orthonormal with respect to the measure ${\rm d}\mu(\lambda)\equiv {\rm d}\lambda\omega(\lambda)\equiv{\rm d}\lambda \exp\{-NV^{(\alpha)}(\lambda)\}$ on the real axis 
\begin{equation}
\int_{-\infty}^\infty {\rm d}\mu(\lambda)P_n(\lambda)P_m(\lambda)=\delta_{mn} \ . 
\end{equation} 

Along with the polynomials $P_n(\lambda)$, we will require the functions $\varphi_n(\lambda)$
\begin{equation}
\varphi_n(\lambda)\equiv P_n(\lambda)\exp\{-\frac{N}{2}V^{(\alpha)}(\lambda) \} \ ,
\label{defvarphi}
\end{equation} 
which are by construction orthonormal on the real axis with respect to the measure ${\rm d}\lambda$.
The kernel can be expressed in terms of the $\varphi_k$ as
\begin{eqnarray}
K_N(\lambda,\lambda') & = & \sum_{k=0}^{N-1}\varphi_k(\lambda)\varphi_k(\lambda') 
\label{kernelUI} \\ 
                      & = & c_N\frac{\varphi_N(\lambda')\varphi_{N-1}(\lambda)-\varphi_N(\lambda)\varphi_{N-1}(\lambda')}{\lambda'-\lambda} \ ,
\label{kernelUII}
\end{eqnarray}
where the latter equality is derived using the Christoffel-Darboux formula \cite{Szego}.
The three term recursion relation
\begin{equation}
\lambda P_{n-1}(\lambda)= c_n P_n(\lambda)- c_{n-1}P_{n-2}(\lambda)
\label{recrel}
\end{equation} 
determines the coefficients $c_n$ of which we have just encountered $c_N$. The aim is to derive a differential equation for the orthonormal functions $\varphi_k(\lambda)$ \cite{KandFWork}. In order to do so, we write the derivative of $P_n(\lambda)$ as
\begin{equation}
\frac{{\rm d}P_n(\lambda)}{{\rm d}\lambda}= A_n(\lambda)P_{n-1}(\lambda)-B_n(\lambda)P_n(\lambda) \ ,
\label{difP}
\end{equation} 
where the functions $A_n(\lambda)$ and $B_n(\lambda)$ are \cite{KandFWork}
\begin{eqnarray}
A_n(\lambda) & = & Nc_n\int_{-\infty}^\infty \frac{{\rm d}\mu(t)}{t-\lambda}\left(\frac{{\rm d}V^{(\alpha)}(t)}{{\rm d}t}-\frac{{\rm d}V^{(\alpha)}(\lambda)}{{\rm d}\lambda}\right)P^2_n(t) \label{Adef}\\
B_n(\lambda) & = & Nc_n\int_{-\infty}^\infty \frac{{\rm d}\mu(t)}{t-\lambda}\left(\frac{{\rm d}V^{(\alpha)}(t)}{{\rm d}t}-\frac{{\rm d}V^{(\alpha)}(\lambda)}{{\rm d}\lambda}\right)P_n(t)P_{n-1}(t) \label{Bdef}\ .
\end{eqnarray}
These expressions for $A_n(\lambda)$ and $B_n(\lambda)$ can be obtained by expressing the left hand side of eqn.\,(\ref{difP}) through an expansion 
\begin{equation}
\frac{{\rm d}P_n(\lambda)}{{\rm d}\lambda}= \int_{-\infty}^\infty {\rm d}\mu(t) \frac{{\rm d}P_n(t)}{{\rm d}t}\sum_{k=0}^{n-1}P_k(\lambda)P_k(t)     \ ,
\end{equation} 
and then rewriting this integral using partial integration and the Christoffel-Darboux formula.

Using eqs.\,(\ref{Adef}), (\ref{Bdef}), and the three term recursion relation eqn.\,(\ref{recrel}), one finds the following identity
\begin{equation}
B_n(\lambda)+B_{n-1}(\lambda)+N\frac{{\rm d}V^{(\alpha)}(\lambda)}{{\rm d} \lambda}=\frac{\lambda}{c_{n-1}}A_{n-1}(\lambda) \ .
\label{ABidentity}
\end{equation} 
We are now able to derive a differential equation for the orthonormal functions $\varphi_n$. Differentiating eqn.\,(\ref{difP}) with respect to $\lambda$ and  making use of the recursion relation eqn.\,(\ref{recrel}) and of the defining eqn.\,(\ref{defvarphi}) for $\varphi_n$, we obtain \cite{KandFWork} 
\begin{equation}
\frac{{\rm d}^2\varphi_n(\lambda)}{{\rm d} \lambda^2}-\mathcal{F}_n(\lambda)\frac{{\rm d}\varphi_n(\lambda)}{{\rm d} \lambda}+\mathcal{G}_n(\lambda)\varphi_n(\lambda)=0
\label{exactdif}
\end{equation}
where
\begin{eqnarray}
\mathcal{F}_n(\lambda) & \equiv & \frac{{\rm d}}{{\rm d}\lambda}{\rm log} A_n(\lambda)   
\label{defF}
\end{eqnarray}
and 
\begin{eqnarray}
\mathcal{G}_n(\lambda) & \equiv & \frac{c_n}{c_{n-1}}A_n(\lambda)A_{n-1}(\lambda)-\left(B_n(\lambda)+\frac{N}{2}\frac{{\rm d} V^{(\alpha)}(\lambda)}{{\rm d}\lambda}\right)^2   \\
 & + & \frac{{\rm d}}{{\rm d}\lambda}\left(B_n(\lambda)+\frac{N}{2}\frac{{\rm d} V^{(\alpha)}(\lambda)}{{\rm d}\lambda}\right) - \frac{{\rm d}{\rm log} \,\, A_n(\lambda)}{{\rm d}\,\lambda} \left(B_n(\lambda)+\frac{N}{2}\frac{{\rm d} V^{(\alpha)}(\lambda)}{{\rm d}\lambda}\right) \ . \nonumber
\label{defG}
\end{eqnarray}
It is straightforward to obtain the kernel, provided that we can solve these equations for $\varphi_N$ and $\varphi_{N-1}$, cf. eqn.\,(\ref{kernelUII}).  
\vspace{2mm}

In accordance with \cite{KandFWork} we choose the following convenient representation of the functions $A_n(\lambda)$ and $B_n(\lambda)$, 
\begin{eqnarray}
A_n(\lambda) & = & A^{(n)}_{\rm reg}(\lambda)+\alpha A^{(n)}_{\rm sing}(\lambda) \\
B_n(\lambda) & = & B^{(n)}_{\rm reg}(\lambda)+\alpha B^{(n)}_{\rm sing}(\lambda) \ ,
\end{eqnarray}  
where the regular parts depend only on the polynomial terms in the general potential, eqn.\,(\ref{genpot}), and the singular part depends on the term $\log [f(\lambda)]^2$. Since we have not changed the polynomial part, we are led to the same expressions for the regular parts of $A_N(\lambda)$ and $B_N(\lambda)$ as obtained in \cite{KandFWork}. As our interest is the microscopic limit, we evaluate the differential equation, eqn.\,(\ref{exactdif}), in this limit. In doing so it is essential to keep track of the $N$-dependence of the various terms. We will assume that the recursion coefficients $c_n$ and the functions $A_{\rm reg}^{(n)}$ and $B_{\rm reg}^{(n)}$ approach smooth functions in the large-$N$ limit so that
\begin{eqnarray}
c_{N\pm1} & = &  c_N+\mathcal{O}(1/N) \\
A_{\rm reg}^{(N\pm1)} & = &  A_{\rm reg}^{(N)}+\mathcal{O}(1/N) \\
B_{\rm reg}^{(N\pm1)} & = &  B_{\rm reg}^{(N)}+\mathcal{O}(1/N)\label{Bbeh} \ .
\end{eqnarray}
With these conditions, the function $A_{\rm reg}^{(N)}$ is related to the macroscopic spectral density \cite{KandFWork,multicritical}
\begin{equation}
\rho(\lambda)=\lim_{N\to\infty}\frac{1}{N\pi}A_{\rm reg}^{(N)}(\lambda)\sqrt{1-(\lambda/a)^2} \ ,
\label{Adens}
\end{equation}
where $a\equiv\lim_{N\to\infty}2c_N$ can be identified with the endpoints of the spectrum. From this, we infer that $A_{\rm reg}^{(N)}(\lambda)$ is of order $N$ in the limit $N\to\infty$, provided that $\lambda$ is sufficiently far from the spectral endponts.

Using the assumption of eqn.\,(\ref{Bbeh}) to evaluate eqn.\,(\ref{ABidentity}) in the large-$N$ limit, we obtain
\begin{equation}
B_N(\lambda)+\frac{N}{2}\frac{{\rm d} V^{(\alpha)}(\lambda)}{{\rm d}\lambda}=\frac{\lambda}{2c_N}A_N(\lambda)-\frac{\alpha}{2}(B_{\rm sing}^{(N-1)}-B_{\rm sing}^{(N)}) \ .
\label{ABidentity-NI}
\end{equation} 
 
% Due to eqn.\,(\ref{kernelUII}) we need only to consider $n=N\gg 1$, in this limit
%\begin{equation}
%A^{(N)}_{\rm reg}(\lambda)=\frac{\pi\rho_D(\lambda)}{\sqrt{1-(\lambda/(2c_N))^2}} \ ,
%\label{Areg}
%\end{equation}
%where the Dyson density $\rho_D$ is independent of $V^{(\alpha)}_{\rm sing}$ (see \cite{KandFWork} their eqn.\,(31)), and
%\begin{equation}
%B^{(N)}_{\rm reg}(\lambda)=\frac{\lambda}{2c_n}A^{(N)}_{\rm reg}(\lambda)-\frac{{\rm d}V}{d \lambda}\Big|_\lambda \ .
%\end{equation}

The singular parts $A^{(N)}_{\rm sing}(\lambda)$ and $B^{(N)}_{\rm sing}(\lambda)$ depend on $f(\lambda)$. We will now prove that the $f$-dependent terms are are smaller than the $f$-independent terms by at least a factor of $1/N$ in the microscopic limit.
\vspace{2mm}

\noindent
First, we determine the $\lambda$ dependence of $A^{(N)}_{\rm sing}(\lambda)$ and $B^{(N)}_{\rm sing}(\lambda)$. The singular part $A^{(N)}_{\rm sing}(\lambda)$ reads
\begin{eqnarray}
A^{(N)}_{\rm sing}(\lambda) & = & -c_N\int_{-\infty}^\infty\frac{{\rm d}\mu(t)}{t-\lambda}\left(\frac{{\rm d}V^{(\alpha)}_{\rm sing}(t)}{{\rm d}t}-\frac{{\rm d}V^{(\alpha)}_{\rm sing}(\lambda)}{{\rm d}\lambda}\right)P^2_N(t) \nonumber \\
 & = & -c_N\int_{-\infty}^\infty\frac{{\rm d}\mu(t)}{t-\lambda}\left(\frac{{\rm d}\log [f(t)]^2}{{\rm d}t}-\frac{{\rm d}\log [f(\lambda)]^2}{{\rm d}\lambda}\right)P^2_N(t)\nonumber \\ 
 & = & -2c_N\int_{-\infty}^\infty\frac{{\rm d}\mu(t)}{t-\lambda}\left(\frac{f'(t)f(\lambda)-f(t)f'(\lambda)}{f(t)f(\lambda)} \right)P^2_N(t)
\ .
\label{AS}
\end{eqnarray}  
We now introduce the defining expansion of $f(\lambda)$ and expand $1/(t-\lambda)$ in $\lambda$: 
\begin{eqnarray}
 A^{(N)}_{\rm sing}(\lambda) 
 & = & \frac{-2c_N}{\lambda+f_3\lambda^3+\ldots}\int_{-\infty}^\infty{\rm d}\mu(t)\frac{1}{tf(t)}(1+\frac{\lambda}{t}-\ldots) \nonumber \\
 & & \hspace{2cm}\cdot\left(f'(t)(\lambda+f_3\lambda^3+\ldots)-f(t)(1+3f_3\lambda^2+\ldots) \right)P^2_N(t) \nonumber\\
 & = & \frac{-2c_N}{\lambda+f_3\lambda^3+\ldots}\int_{-\infty}^\infty{\rm d}\mu(t)\left\{-\frac{1}{t}-\frac{\lambda}{t^2}+\frac{\lambda}{t}\frac{f'(t)}{f(t)}+\ldots\right\}P^2_N(t) \nonumber\\
 & = & \frac{-2c_N}{1+f_3\lambda^2+\ldots}\int_{-\infty}^\infty{\rm d}\mu(t)\left\{-\frac{1}{t^2}+\frac{1}{t}\frac{f'(t)}{f(t)}+\ldots \right\}P^2_N(t)
\label{evaluationI}
\end{eqnarray}
In the final equality we have used the fact that both the weight $\omega(t)$ of the measure and $P^2_N(t)$ are even in $t$. $P^2_N(t)$ is even in $t$ since $P_n(t)$ has parity  $P_n(-t)=(-1)^n P_n(t)$, as can be shown using the three term recursion relation eqn.\,(\ref{recrel}) iteratively.
\vspace{2mm}

The singular part of $B_N(\lambda)$ is evaluated in an almost analogous fashion. The only extra ingredient is that one must use the recursion relation eqn.\,(\ref{recrel}) in order to perform the integrations. We find that the term proportional to $1/\lambda$ is independent of $f$. In addition to this term, which is identical to the one obtained in \cite{KandFWork}, we obtain a constant term and terms proportional to positive powers of $\lambda$. Denoting the sum of the additional terms by $T_B(N,\lambda)$, we have
\begin{equation}
B_{\rm sing}^{(N)}(\lambda)=\frac{1-(-1)^N}{\lambda}+T_B(N,\lambda) \ . 
\label{Bsing}
\end{equation}

\noindent
To complete the proof, we determine the differential equation in the microscopic limit.
Inserting $B_{\rm sing}^{(N)}(\lambda)$ into eqn.\,(\ref{ABidentity-NI}), we find for the large-$N$ limit of eqn.\,(\ref{ABidentity})
\begin{equation}
\!\!\!\!\!\!\!\! B_N(\lambda)+\frac{N}{2}\frac{{\rm d} V^{(\alpha)}(\lambda)}{{\rm d}\lambda}=\frac{\lambda}{2c_N}A_N(\lambda)-(-1)^N\frac{\alpha}{\lambda}-\frac{\alpha}{2}(T_B(N-1,\lambda)-T_B(N,\lambda)) \ .
\label{ABidentity-NII}
\end{equation} 
Inserting this into the defining eqn.\,(\ref{defG}) for $\mathcal{G}_n$, we can determine its asymptotic behaviour for large $N$
\begin{eqnarray}
\mathcal{G}_N(\lambda) & = & A_N^2(\lambda)(1-\frac{\lambda^2}{a^2})+\frac{(-1)^N\alpha-\alpha^2}{\lambda^2}+(-1)^N\frac{\alpha}{\lambda}\frac{A'_N(\lambda)}{A_N(\lambda)} \\ 
 & & +\frac{A_N(\lambda)}{c_N}(\frac{1}{2}+(-1)^N\alpha) 
-\frac{\alpha^2}{4}(T_B(N-1,\lambda)-T_B(N,\lambda))^2 \nonumber \\ 
 & & +\frac{\alpha}{2}(T_B(N-1,\lambda)-T_B(N,\lambda))\left(\frac{\lambda}{c_N}A_N(\lambda)-2(-1)^N\frac{\alpha}{\lambda}+\frac{A'_N(\lambda)}{A_N(\lambda)}\right)\nonumber\\
 & & -\frac{\alpha}{2}\frac{\rm d}{{\rm d} \lambda}(T_B(N-1,\lambda)-T_B(N,\lambda))  \ . \nonumber
\end{eqnarray} 

In the microscopic limit, two terms in $\mathcal{G}_N(\lambda)$ are dominant (i.e., of order $N^2$). To see this we consider the $N$ dependence in the microscopic limit of the various terms involved. As noted, eqn.\,(\ref{Adens}) implies that
\begin{equation}
A^{(N)}_{\rm reg}(\lambda) \propto N \ .
\end{equation} 
From the evaluation of the $\lambda$-dependence of $A^{(N)}_{\rm sing}(\lambda)$ and $B^{(N)}_{\rm sing}(\lambda)$ eqn.\,(\ref{evaluationI}) and eqn.\,(\ref{Bsing}), we have
\begin{eqnarray}
A^{(N)}_{\rm sing}(\lambda) & \propto & \mathcal{O}(1) \label{Asingpropto}\\
B^{(N)}_{\rm sing}(\lambda) & \propto & (1-(-1)^N)N+\mathcal{O}(1) \ .
\end{eqnarray}
Finally, the macroscopic density and, hence, $A^{(N)}_{\rm reg}$ through eqn.\,(\ref{Adens}) is independent of $\lambda$ for $\lambda\sim0$. We can thus use eqn.\,(\ref{evaluationI}) to obtain
\begin{equation}
A'_N(\lambda)  \propto \mathcal{O}(1) \ .
\end{equation} 
Keeping only terms of order $N^2$ in eqn.\,(\ref{exactdif}) results in 
\begin{eqnarray}
\frac{{\rm d}^2\varphi_n(x)}{{\rm d} x^2} 
 & + & \left\{\frac{(A^{(N)}_{\rm reg}(0))^2}{N^2}+\frac{(-1)^N\alpha-\alpha^2}{x^2}   \right\}\varphi_n(x)=0 \ .
\label{approkdiffI}
\end{eqnarray}
Here we have changed variables to $x=\lambda N$ according to the microscopic scaling limit and divided by $N^2$.
The important feature is that there are no references to the expansion coefficients $f_k$ in eqn.\,(\ref{fexp}). Making use of eqn.\,(\ref{kernelUII}), the kernel of the UE follows cf. \cite{KandFWork}.
Since every orthonormal functions which enters in this kernel is universal with respect to deformations in both the potential $V$ and logarithm $f$ we infer that the microscopic correlators obtained from this kernel by eqn.\,(\ref{rhofraK}) are universal. This completes the proof.
\vspace{4mm}

The restriction to an odd polynomial in eqn.\,(\ref{fexp}) was necessary because the weight in the measure need to be even and the lowest order term in eqn.\,(\ref{fexp}) has to be linear. Following the lines of the proof given above, it is simple to prove that: If one replaces $f(\lambda_i)$ in $\mathcal{Z}_{\rm UE}$ with $f(|\lambda_i|)$ then without affecting the microscopic spectral correlators one can introduce even powers in the definition of $f$. That is, microscopic spectral correlators of the partition function
\begin{eqnarray}
\mathcal{Z} & \sim &  \int_{-\infty}^\infty \prod_{i=1}^N\left({\rm d}\lambda_i [f(|\lambda_i|)]^{2\alpha} \exp\{-NV(\lambda_i)\}\right) |\Delta(\lambda)|^2 \ .
\label{invUZ(l)II}
\end{eqnarray}
are independent of the coefficients $f_k\in\mathbb{R}$ of the polynomial 
\begin{equation}
f(\lambda)=\sum^\infty_{k=1}f_{k}\lambda^{k}
\label{fexpII}
\end{equation}
with $f_1\not=0$.

While this partition function to our knowledge does not have any explicit matrix analogue it will nevertheless prove useful as we turn to the chiral unitary ensemble.

%%%%%%%%%%%%%%%%%%%

\section{Logarithmic universality in the chiral unitary ensemble}

 Chiral unitary random matrix ensembles were originally introduced \cite{ShuVer,VerZah,Verorig} to model the consequences of chiral symmetry in the spectrum of the Euclidean Dirac operator of QCD. The structure of the random matrix representation, $M$, of the Dirac operator was determined by the vanishing anticommutator of $M$ and $\gamma_5$. This anticommutation has one particularly simple spectral consequence: The eigenvalues come in pairs of opposite sign. In extending the universality of the chiral ensemble, we shall again replace the matrix, $M$, in the determinant by an expansion in powers of the matrix. However, we must be somewhat careful because we will need to retain a remnant of chiral symmetry. Here we show that the condition for universality of the microscopic spectral correlators is complex conjugation symmetry of the spectrum and {\sl not} chiral symmetry as established through the anticommutation with $\gamma_5$.

To this end, we we will prowe the assertion that:{\sl The microscopic correlators of the partition function
\begin{equation}
\mathcal{Z}_{\chi{\rm UE}} = \int {\rm d}M {\rm det}^{\tilde{\alpha}}\left( \tilde{f}(M) \right)\exp\{-N{\rm Tr}\tilde{V}(WW^\dagger)\} \ ,
\label{invZ(M)}  
\end{equation}
where $\tilde\alpha$ is an integer, are independent of the coefficients $\tilde{f}_k\in\mathbb{R}$ of the polynomial 
\begin{equation}
\tilde{f}(M)=\sum^\infty_{k=1}\tilde{f}_kM^k
\label{tildefexp}
\end{equation}
when $\tilde{f}_1\not=0$.}
\vspace{4mm}

By definition the hermitian $2N\times 2N$ matrix $M$ is block off-diagonal:
\begin{equation}
M\equiv\left(\begin{array}{cc}
              0 & iW \\ iW^\dagger & 0
             \end{array}\right) \ ,
\label{M}
\end{equation}
where $W$ is a complex $N\times N$ matrix. The non-singular potential is given by
\begin{equation}
\tilde{V}(WW^\dagger)=\sum_{k=1}^p\frac{\tilde{g}_{k}}{k}(WW^\dagger)^k\ , \ \ \tilde{g}_k\in \mathbb{R} \ {\rm with}\ \tilde{g}_p>0 \ .
\end{equation}
Again, it is sufficient to concider $\tilde{f}_1=1$.

\subsection{Proof of logarithmic universality in the $\chi$UE}

The strategy of the proof is first to write the $\chi$UE partition function in an eigenvalue representation and then to relate this partition function to that of the eqn.\,(\ref{invUZ(l)II}) introducing appropriate shifts of the arguments. This will allow us to express the kernel of the $\chi$UE in terms of the orthonormal functions of the UE and, hence, carry over the logarithmic universality of the UE to the $\chi$UE.
\vspace{3mm}

We rewrite the $M$ integration in the generating functional in angular coordinates $W= V\Lambda U^\dagger$, where $V$ and $U$ are unitary matrices and $\Lambda$ is a diagonal matrix with real and positive entries $\lambda_i$, $i=1,\ldots,N$. This is useful since $\mathcal{Z}_{\chi{\rm UE}}$ is invariant under the transformation of $W\to U^\dagger WV$. This invariance is obvious because the transformation of $M$ and hence of $\tilde{f}(M)$ is unitary.
Under the unitary transformation of $M$, the measure ${\rm d}M$ changes to \cite{morris}
\begin{equation}
{\rm d}M=\prod_{i=1}^N{\rm d}\lambda_i^2[{\rm d}V][{\rm d}U]|\Delta(\lambda^2)|^2 \ , \ {\rm where} \ \Delta(\lambda^2)=\prod_{i<j}(\lambda_i^2-\lambda_j^2) \ .
\label{changemeasure}
\end{equation}
The eigenvalues $\lambda^2$ of $WW^\dagger$ are related to the eigenvalues of $M$, which are $\pm i\lambda$.

It follows from the reflection symmetry of the spectrum of $M$ and the choice of real coefficients $\tilde{f}_k$ that the spectrum of $\tilde{f}(M)$ has complex conjugation symmetry. The determinant factor can therefore be written
\begin{equation}
{\rm det}\left( \tilde{f}(M) \right)=\prod_{j=1}^N \tilde{f}(i\lambda_j)\tilde{f}(i\lambda_j)^*.
\label{49}
\end{equation}
Eqn.\,(\ref{49}) show that the determinant is a function of the $\lambda_j^2$.
Thus, the complex conjugation symmetry of the spectrum of $\tilde{f}(M)$ is the remnant of chiral symmetry which we have retained to ensure that the determinant can be written in terms of the eigenvalues of $WW^\dagger$.

Due to the cyclic property of the determinant and the trace, the integrations over $V$ and $U$ become trivial, producing irrelevant factors in $\mathcal{Z}_{\chi{\rm UE}}$. The partition function now reads 
\begin{eqnarray}
\mathcal{Z}_{\chi{\rm UE}} & \sim   & \int_0^{\infty} \prod_{j=1}^N\left({\rm d}\lambda_j^2 |\tilde{f}(i\lambda_j)|^{2 \tilde{\alpha}} \exp\{-N\tilde{V}(\lambda_j^2)\}\right) |\Delta(\lambda^2)|^2 \\ 
& \sim   & \int_{-\infty}^\infty \prod_{j=1}^N\left({\rm d}z_j |z_j||\tilde{f}(iz_j)|^{2 \tilde{\alpha}} \exp\{-N\tilde{V}(z_j^2)\}\right) |\Delta(z^2)|^2 \ .
\end{eqnarray}

This partition function is not of the same form as that of the unitary ensemble eqn.\,(\ref{invUZ(l)}). However, inspired by \cite{multicritical}, we can carry over the universality argument of the preceding section. 
Given the set of coefficients $\tilde{f}_k$ defining the polynomial $\tilde{f}$, we introduce polynomials $\tilde{P}^{(\tilde{f}, \tilde{\alpha})}_l(z^2)$ orthonormal with respect to the measure ${\rm d}z|z||\tilde{f}(iz)|^{2 \tilde{\alpha}} \exp\{- N\tilde{V}(z^2)\}$ on the real axis. Notice that it is possible to replace $|\Delta(z^2)|^2$ by a determinant involving only the even polynomials $\tilde{P}^{(\tilde{f}, \tilde{\alpha})}_l(z^2)$.
The idea is to identify the polynomials $\tilde{P}^{(\tilde{f}, \tilde{\alpha})}_l(z^2)$ with the polynomials $P^{(f,\alpha)}_{2l}(z)$ for the partition function eqn.\,(\ref{invUZ(l)II}).
%\begin{equation}
%\tilde{P}^{(\tilde{f}, \tilde{\alpha})}_l(z^2)\equiv P^{(f(\tilde{f}),\tilde{\alpha}+1/2)}_{2l}(z)
%\end{equation}
We will now show that this is justified if $\alpha=\tilde{\alpha}+1/2$ and the expansion coefficients of $f$ in eqn.\,(\ref{fexpII}) are related to those of $\tilde{f}$ in eqn.\,(\ref{tildefexp}). This can be done provided that we identify
\begin{equation}
\tilde{V}(z^2)\equiv 2V(z) \ . 
\label{Vinden} 
\end{equation}
The proof is then completed by expressing the kernel of the $\chi$UE in terms of the orthonormal functions $\varphi_l$ of the partition function eqn.\,(\ref{invUZ(l)II}). Since these orthonormal functions are universal in the microscopic limit, the kernel is also universal in the microscopic limit.
\vspace{2mm}
  
In order to find the set of coefficients $\{f_k\}$ in eqn.\,(\ref{fexpII}) such that $|z||\tilde{f}(iz)|^{2 \tilde{\alpha}}=f(|z|)^{2(\tilde{\alpha}+1/2)}$ is satisfied for a given set $\{\tilde{f}\}$, let us first expand the left hand side of this identity. Using the fact that $\tilde{f}_1=1$, we find
\begin{eqnarray}
 & & |z||\tilde{f}(iz)|^{2 \tilde{\alpha}}\nonumber\\
 & = & |z|\left\{ \left(\sum^\infty_{k=0}\tilde{f}_{2k+1}(-1)^{k}z^{2k+1}\right)^2 +\left(\sum^\infty_{k=0}(-1)^{k+1}\tilde{f}_{2k+2}z^{2k+2}\right)^2\right\}^{\tilde{\alpha}} \nonumber\\
 & = &  |z|\left\{ \sum^\infty_{k,k'=0}\left[\tilde{f}_{2k+1}\tilde{f}_{2k'+1}(-1)^{k+k'}z^{2(k+k')+2}+\tilde{f}_{2k+2}\tilde{f}_{2k'+2}(-1)^{k+k'+2}z^{2(k+k')+4}\right]\right\}^{\tilde{\alpha}}\nonumber \\
 & = &  \!\!\!\!\!\!\!\!\!\!\!\sum^\infty_{ \ \ \ k_1,k_1',\ldots,k_{\tilde{\alpha}},k_{\tilde{\alpha}}'=0}\!\!\!\!\!\!\!\!\!\!\!\!\!\!(-1)^{\sum_{i=1}^{\tilde{\alpha}} k_i+k_i'}\prod_{j=1}^{\tilde{\alpha}}\left[\tilde{f}_{2k_j+1}\tilde{f}_{2k_j'+1}+\tilde{f}_{2k_j+2}\tilde{f}_{2k_j'+2}z^2\right]|z|^{2(\sum_{i=1}^{\tilde{\alpha}} k_i+k_i')+2\tilde{\alpha}+1} \nonumber\\
 & = & |z|^{2(\tilde{\alpha}+\frac{1}{2})}+\ldots \ .
\label{chUexp}
\end{eqnarray}
Since only odd powers of $|z|$ appear, it is  appropriate to compare this expansion with that of $f(|z|)^{2(\tilde{\alpha}+1/2)}$ of eqn.\,(\ref{fexpII}) where $f_{2k+2}=0$.  
\begin{eqnarray}
[f(|z|)]^{2(\tilde{\alpha}+\frac{1}{2})} & = & \left(\sum^\infty_{k=0}f_{2k+1}|z|^{2k+1}\right)^{2(\tilde{\alpha}+\frac{1}{2})}\nonumber \\
 & = &    \!\!\!\!\!\!\sum^\infty_{ \ \ \ k_0,k_1,k_1',\ldots,k_{\tilde{\alpha}},k_{\tilde{\alpha}}'=0}\!\!\!\!\!\!\!\!\!\! f_{2k_0+1}\prod_{j=1}^{\tilde{\alpha}} f_{2k_j+1}f_{2k_j'+1}|z|^{2(k_0+\sum_{i=1}^{\tilde{\alpha}} k_i+k_i')+2\tilde{\alpha}+1} \nonumber \\
 & = & |z|^{2(\alpha+\frac{1}{2})}+\cdots \ .
\label{Uexp}
\end{eqnarray}
The proof that it is possible to choose the $f_{2k+1}$ for an arbitrary choice of coefficients $\tilde{f}_{k}$ such that the expansions of eqn.\,(\ref{Uexp}) and eqn.\,(\ref{chUexp}) are identical to all orders can be made iteratively: First, note that the lowest-order terms in the two expansions are identical. Assume that the identity holds for each term up to power $2n+2\alpha+1$ of $|z|$. Now, the next term in the $\chi$UE expansion yields some coefficient for $|z|^{2(n+1)+2\alpha+1}$, and this coefficient has to be matched by
\begin{displaymath}
\sum_{k_0+k_1+k_1'+\ldots+k_{\tilde{\alpha}}+k_{\tilde{\alpha}}'= n+1,\ k_i\in \mathbb{N}_0}\!\!\!\!\!f_{2k_0+1}\prod_{j=1}^{\tilde{\alpha}} f_{2k_j+1}f_{2k_j'+1} \ . 
\end{displaymath}
Since $f_1=1$, we find that $2\tilde{\alpha}+1$ of the terms in this summation are $f_{2(n+1)+1}$. As $f_{2(n+1)+1}$ does not appear in any lower order term, one can choose it to and match the $\chi$UE expansion of eqn.\,(\ref{chUexp}) and the UE expansion of eqn.\,(\ref{Uexp}) to order $2n+2\alpha+1$ in $|z|$. This completes the proof by induction.

%If the $f_k$, $k\geq 2$, are given by the $\tilde{f}_k$, $k\geq 2$, by the expansion eqn.\,(\ref{chUexp}) one obtain the polynomials sought. In the microscopic limit these polynomials therefore again show the universal behaviour.
\vspace{2mm}

In order to determine the kernel of the $\chi$UE, we introduce orthonormal functions $\tilde{\varphi}_l$ as in eqn.\,(\ref{defvarphi})
\begin{equation}
\tilde{\varphi}_l^{(\tilde{f}, \tilde{\alpha})}(z^2)\equiv |z|^{\frac{1}{2}}|\tilde{f}(iz)|^ {\tilde{\alpha}}\exp\{-\frac{N}{2}\tilde{V}(z^2)\}\tilde{P}^{(\tilde{f}, \tilde{\alpha})}_l(z^2).
\end{equation}

Using the relation between the potentials imposed in eqn.\,(\ref{Vinden}), we see that 
\begin{equation}
\tilde{\varphi}_l^{(\tilde{f},\tilde{\alpha})}(z^2) =\varphi_{2l}^{(f(\tilde{f}),\tilde{\alpha}+1/2)}(z) \ . 
\end{equation}
The kernel is obtained by evaluating
\begin{equation}
K_N(z^2,w^2) = \sum_{l=0}^{N-1}\tilde{\varphi}_l^{(\tilde{f},\tilde{\alpha})}(z^2)\tilde{\varphi}_l^{(\tilde{f},\tilde{\alpha})}(w^2) \ .
\end{equation}
Following the derivation in section 3 of \cite{multicritical}, i.e., using the Christoffel-Draboux formula, we find 
\begin{equation}
K_N(z^2,w^2)=c_{2N}\frac{w\varphi^{(f(\tilde{f}),\tilde{\alpha}+\frac{1}{2})}_{2N}(z)\varphi^{(f(\tilde{f}),\tilde{\alpha}+\frac{1}{2})}_{2N-1}(w)-z\varphi^{(f(\tilde{f}),\tilde{\alpha}+\frac{1}{2})}_{2N-1}(z)\varphi^{(f(\tilde{f}),\tilde{\alpha}+\frac{1}{2})}_{2N}(w)}{z^2-w^2} \ .
\end{equation}

It is immediately seen that the kernel is universal in the microscopic limit since the orthonormal functions $\varphi_l$ are independent of both $\tilde{V}$ and $\tilde{f}$ in that limit. 

\section{Example: The Ginsparg-Wilson relation}

As an indication of the utility of the results obtained above, we consider the partition function eqn.\,(\ref{invZ(M)}) with the choice
\begin{equation}
D\equiv\tilde{f}(M)=\frac{1}{a}[1-e^{aM}]=-\sum_{k=1}^\infty \frac{M^k}{k!}a^{k-1} \ ,
\end{equation}
where $M$ is given by eqn.\,(\ref{M}) and $a$ is a real number. This choice falls within the scope of the results in section 3 since the expansion coefficients are real. Hence, the microscopic spectral correlators in that case are identical to those obtained for $\tilde{f}(M)=M$ which is used when considering QCD in 4 dimensions with $\chi$UE \cite{ShuVer,VerZah,Verorig}. That model contains $SU(\tilde{\alpha})\times SU(\tilde{\alpha})$ chiral symmetry since 
\begin{equation}
\{M,\gamma_5\}=0 \ ,
\label{anticom}
\end{equation}
where
\begin{equation}
\gamma_5={\rm diag}(1,1,\ldots,1,-1,-1,\ldots,-1) \ .
\end{equation}
It is readily seen that $D^\dagger=\gamma_5D\gamma_5$ and that the eigenvalues of $D$ lie on a circle in the complex plane. Further, using eqn.\,(\ref{anticom}) it is easy to show that
\begin{equation}
\{D,\gamma_5\}=aD\gamma_5D \ .
\end{equation}
This anti-commutation relation is known as the Ginsparg-Wilson relation \cite{GW} and implies a $SU(\tilde{\alpha})\times SU(\tilde{\alpha})$ symmetry of the model \cite{luscherII}. This example shows that genuine chiral symmetry and the Ginsparg-Wilson-L\"uscher symmetry lead to identical microscopic correlators. The Ginsparg-Wilson relation is of relevance when constructing lattice actions without fermion doubling, see, e.g., \cite{relevant}. Regarding $a$ as the lattice spacing, it is clear that the Ginsparg-Wilson-L\"uscher symmetry will reduce to genuine chiral symmetry in the limit $a\to 0$. The present argument show however that the microscopic spectral correlators will be obtained for {\sl all} $a$ and that, in this regard, there is nothing special about the $a\to 0$ limit. The first study of random matrix correlations in a Ginsparg-Wilson symmetric lattice action have appeared recently \cite{CLBlatt98}.

\section{Summary}

In this paper we have extended the universality class of the unitary invariant random matrix ensemble and of the chiral unitarily invariant random matrix ensemble to include logarithmic universality. Logarithmic universality expresses independence with respect to deformations of the logarithmic singular part of the general potential.  In the case of the $\chi$UE, this means that the original matrix structure in the determinant can be deformed by means of a series expansion in the original block matrix without altering the microscopic spectral behaviour provided that the spectrum associated with this series has a complex conjugation symmetry. This result was used to demonstrate that the matrix appearing in the determinant of the $\chi$UE partition function can be chosen to satisfy the Ginsparg-Wilson relation without altering the microscopic spectral correlators. This provides an indication that that lattice gauge simmulations can satisfy chiral random matrix statistics for finite lattice spacings even when genuine chiral symmetry has been sacrificed in the intrest of eliminating fermion doubling. 
\vspace{3mm}

{\bf \large \noindent Acknowledgements:} The author gratefully acknowledge discussions with P. H. Damgaard, A. Andersen, J. Christiansen and A. D. Jackson.

\end{document}